\documentclass[aps,prl,preprint,amsmath,amssymb,groupedaddress]{revtex4-1}
\usepackage{graphicx}% Include figure files
\usepackage{dcolumn}% Align table columns on decimal point
\usepackage{bm}% bold math
\pdfoutput=1

\begin{document}
	%\preprint{}
	

\title{Room-temperature 1.54~$\mu$m photoluminescence of Er:O$_x$ centers at extremely low concentration in silicon}

\author{Michele Celebrano}
\author{Lavinia Ghirardini}
\author{Paolo Biagioni}
\author{Marco Finazzi}
\affiliation{Dipartimento di Fisica, Politecnico di Milano, Piazza Leonardo da Vinci 32, I--20133 Milano, Italy}
\author{Yasuo Shimizu}
\author{Yuan Tu}
\author{Koji Inoue}
\author{Yasuyoshi Nagai}
\affiliation{The Oarai Center, Institute for Materials Research, Tohoku University, 2145-2 Narita, Oarai, Ibaraki 311-1313, Japan}
\author{Takahiro Shinada}
\affiliation{Center for Innovative Integrated Electronic Systems,Tohoku University, 468-1 Aoba, Aramakiaza-aoba, Aoba, Sendai, Miyagi 980-0845, Japan}
\author{Yuki Chiba}
\author{Ayman Abdelghafar}
\author{Maasa Yano}
\author{Takashi Tanii}
\affiliation{School of Science and Engineering, Waseda University, 3-4-1 Ohkubo, Shinjuku, Tokyo 169-8555, Japan}
\email{tanii@waseda.jp}
\author{Enrico Prati}
\affiliation{Istituto di Fotonica e Nanotecnologie, Consiglio Nazionale delle Ricerche, Piazza Leonardo da Vinci 32, I--20133 Milano, Italy}
\email{enrico.prati@cnr.it}

\date{Compliled \today}

%\ociscodes{(060.2410) Fibers, erbium; (020.1335) Atom optics; (180.1790) Confocal microscopy; (250.5230) Photoluminescence.}

%\doi{\url{http://dx.doi.org/10.1364/ao.XX.XXXXXX}}

\begin{abstract}
The demand for single photon sources at $\lambda$~=~1.54~$\mu$m, which follows from the consistent development of quantum networks based on commercial optical fibers, makes Er:O$_x$ centers in Si still a viable resource thanks to the optical transition of Er$^{3+}$: $^4$I$_{13/2}$ $\rightarrow$ $^4$I$_{15/2}$. Yet, to date, the implementation of such system remains hindered by its extremely low emission rate. In this Letter, we explore the room-temperature photoluminescence (PL) at the telecomm wavelength of very low implantation doses of Er:O$_x$ in Si. The emitted photons, excited by a $\lambda$~=~792~nm laser in both large areas and confined dots of diameter down to 5~$\mu$m, are collected by an inverted confocal microscope. The lower-bound number of detectable emission centers within our diffraction-limited illumination spot is estimated to be down to about 10$^4$, corresponding to an emission rate per individual ion of about 4 $\times 10^{3}$ photons/s.
\end{abstract}

%\setboolean{displaycopyright}{true}

\maketitle
%\thispagestyle{fancy}

%\ifthenelse{\boolean{shortarticle}}{\ifthenelse{\boolean{singlecolumn}}{\abscontentformatted}{\abscontent}}{}

In the last three decades, considerable efforts have been devoted towards the realization of optical gain in erbium doped silicon for its potential capability of bridging silicon CMOS technology with photonics. One of the technological issues consists in the difficulty of achieving strong doping for erbium atoms in the optically active state \cite{Favennec1990}. High implantation doses exceeding ${1.3~\times 10^{18}}$~cm$^{-3}$ induce a linear growth in the concentration of erbium precipitates \cite{Ref5}, and several methods have been attempted to overcome such limitation, including recrystalization of preliminarily amorphized silicon \cite{Ref2,Ref6}, growth of erbium-doped silicon films by molecular beam epitaxy (MBE) \cite{Ref7}, and recoil implantation \cite{Ref8}. Recently, high luminescence efficiency and energy transfer have been demonstrated from Er ions implanted in SiO$_x$ films when co-doped with either gold \cite{Giova2006} or SnO$_2$ nanocrystals \cite{Zhang2014}, as well as in amorphous Si nanodots \cite{lu2016high} and in silicon nanocrystals \cite{acsphotonics16}. With the aim of enhancing erbium luminescence various silicon microcavity geometries have been succesfully doped at relatively high Er concentration \cite{Wang2016, Kalkman2006, Kalkman2006_2}, also achieving laser action.  
On the contrary, ErO$_x$ complex in silicon itself provides a natural solution for creating single photon emission centers by exploiting its low density regime. Its emission wavelength naturally matches optical fiber-based secure quantum networks. Super-enhancement of PL of Er:O complex in silicon has been recently shown at the ratio around 1:1 in high doping regime \cite{SciRep16}, so its exploration at low dose should grant both efficient PL and countable photon regime.

We have already explored low doping of both electronic \cite{shinada2005enhancing,prati2012anderson,prati2016band,prati2015single} and optoelectronic devices \cite{tamura2014array} in the limit of single atom doping by single ion implantation method, as well as deterministic doping by directed self-assembly and spike annealing \cite{shimizu2014behavior}. Low doping of optical species such as Er has been reported only for 1.54~$\mu$m absorption by a transistor operated at cryogenic temperature \cite{Rogge13}, and Pr-doped YAG nanocrystal \cite{kolesov2012optical} at a different wavelength.
With the perspective of achieving controlled single photon emission regime by ErO$_x$ complexes in silicon, here we report on the fabrication and the PL characterization of shallow implanted Si:ErO$_x$.

We systematically study the response to a pump radiation at 792~nm in the mW range by varying the concentration of both oxigen and erbium ions implanted at 20 keV from 10$^{12}$ to $10^{14}$~ions/cm$^{2}$. Room temperature PL is detected both from large square patches and from disks with a radius down to 5~$\mu$m, which contain only about 10$^4$ implanted ions, thus providing a limiting reference for minimum Er atoms required to emit a detectable signal at room temperature if no additional manufacturing such as nanocavities or waveguide fabrication is employed.
Our experimental realization allowed detecting PL from as few as $\approx 10^4$ Er ions with an emission rate of about ${4 \times 10^{3}}$ photons per second per ion.

Frame-like micro patterns for the alignment of Er ion implantation were fabricated on silicon substrates by electron beam lithography (EBL) and wet etching. Samples were prepared from boron-doped (15~Ohm$\cdot$cm) Cz-Si(100) with an oxygen content of ${4\times 10^{17}}$~cm$^{-3}$, which inhibits PL at room temperature \cite{michel1991impurity,priolo1995erbium}.
The Si surface was covered with a thermally grown SiO$_2$ layer with a thickness of 5~nm.
The frame-like square micro patterns had an area of ${100~\mu\mathrm{m}\times 100~\mu}$m and a line width of $10~\mu$m, as depicted in the sketch in Fig.~\ref{fig:Fig2}a. The surface was coated with polymethylmethacrilate (PMMA) resist film of 200~nm thickness, and only the resist film inside the rectangular area was removed by EBL.
Er$^{2+}$ ions were implanted inside the squared area at an energy of 20 keV using an ion implanter equipped with AuSiEr liquid metal ion source and an ${\mathbf{E} \times \mathbf{B}}$ mass filter. The Er dose was varied from ${1 \times 10^{12}}$~cm$^{-2}$ to ${1 \times 10^{14}}$~cm$^{-2}$.
After removing the PMMA film with O$_2$ plasma, the substrates were coated again with PMMA resist film of 100~nm thickness which was used to adjust the implantation depth of O ions relatively to Er ions. 
Then O$^+$ ions were implanted at an energy of 25~keV using a conventional ion implanter equipped with a CO$_2$ gas source.
The implantation dose was altered from ${2.5 \times 10^{12}}$~cm$^{-2}$ to ${1 \times 10^{14}}$~cm$^{-2}$ for each substrate.
After removing the PMMA film, the substrates were annealed at $900~^\circ$C for 30~min in dry N$_2$ atmosphere.
Finally, the SiO$_2$ layer on each substrate was removed in buffered hydrofluoric acid.

\begin{figure}[t!]
	\centering
	\mbox{\includegraphics[width=10 cm]{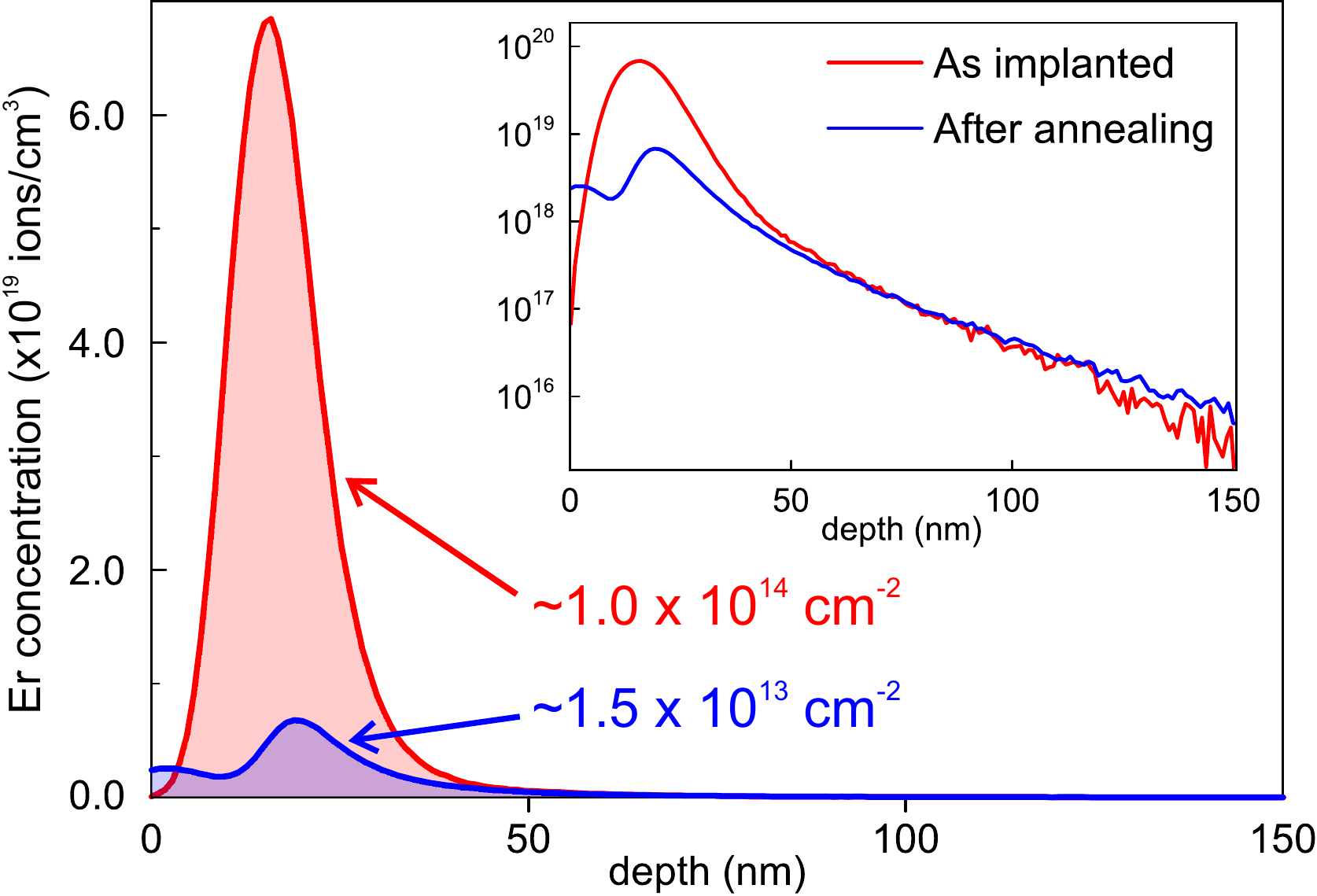}}
	\caption{SIMS depth profiles of a sample with nominal Er concentration of ${\approx 1 \times 10^{14}}$~cm$^{-2}$ before and after annealing. The Er content is evaluated by integrating the concentration in the volume for the whole sample depth (see semi-transparent areas below the curves). The concentration measured after (marked blue) annealing is about 7 times lower than the one recorded before (marked red). Inset: depth profiles plotted in logarithmic scale.}
	\label{fig:Fig1}
\end{figure}

The annealing process is expected to promote an out-diffusion of the Er$^{2+}$ ions from the Si-air interface, hence effectively reducing the final concentration of emission centers contained in Si. To establish the effective Er$^{2+}$ concentration, we performed secondary ion mass spectrometry (SIMS) on a sample featuring the highest nominal implantation level (${1 \times 10^{14}}$~ions/cm$^{2}$) before and after the annealing process, which is identical to the ones devised for the optical characterization. SIMS measurements were performed by using primary ions of O$_{2}^{+}$ at an acceleration energy of 3.0~kV with CAMECA IMS-7f. The detection area was 30~$\mu$m$^{2}$ in diameter. The ion concentration level retrieved by SIMS before annealing matches perfectly the nominal values, confirming the excellent reliability of the implantation technique. On the other hand, as expected, the effective ion concentration after annealing drops to ${1.5 \times 10^{13}}$~cm$^{-2}$ (see Fig.~\ref{fig:Fig1}). Therefore, to account for
such an additional phenomenon, the nominal ion concentration has to be scaled by about a factor 7 in order to attain the effective ion concentration in Si. 
A careful observation of Fig.~\ref{fig:Fig1} allows identifying a pile-up of the Er$^{2+}$ ions close to the Si-air interface in the annealed sample, which we attribute to surface contamination. It is worth noting that the O$^{+}$ amount is also reduced by about a factor 10, as can be derived by combining SRIM ion range simulations and diffusion equation, so the ratio around 1:1 is maintained. At these co-doping regimes, i.e. O:Er~$\leq$~2, density functional computations \cite{raffa2002equilibrium} indicates that the tetrahedral substitutional site is the most stable. In addition to that, according to Ref.~\cite{SciRep16}, this is also the most optically active configuration due to a more favorable crystal field splitting in the case of the singly oxygen coordinated Er in a tetrahedral substitutional site instead of interstitial site.

\begin{figure}[t!]
	\centering
	\mbox{\includegraphics[width=10 cm]{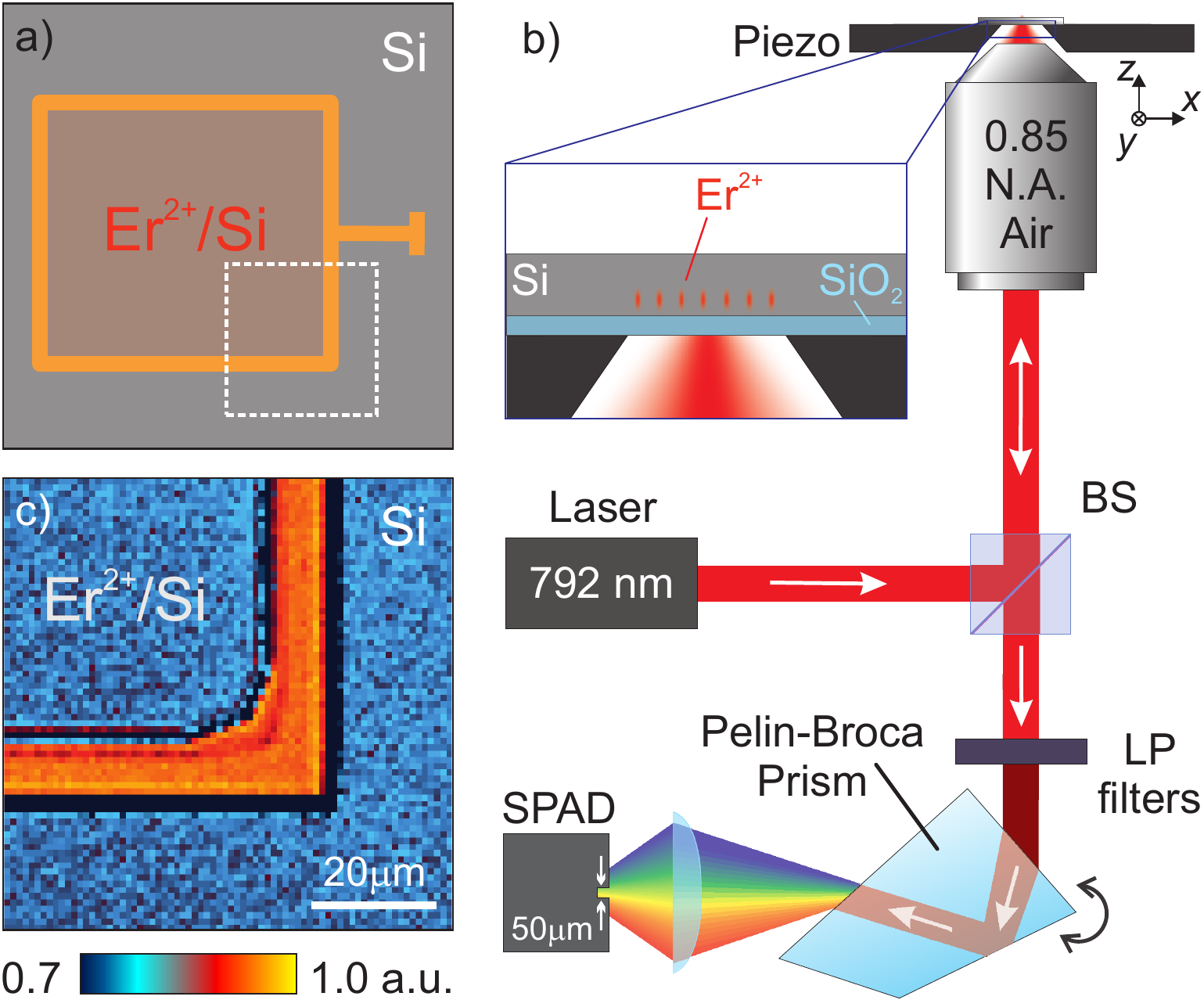}}
	\caption{a) Sketch of an implanted region, identified by gold trances for an easy identification. b) Experimental setup. BS: 70/30 beamsplitter, LP filters: two long-pass filters with 1300 and 1350~nm cut-off wavelengths, respectively; SPAD: InGasAs/InP single photon avalanche diode. Inset: side-view detail of the investigated sample showing the Er$^{2+}$ implanted areas. c) Reflection map recorded at the excitation wavelength (792~nm).}
	\label{fig:Fig2}
\end{figure}

We addressed Er ions implanted in Si and investigated their photoemission properties by means of a customized inverted confocal microscope coupled to a spectrometer (see Fig.~\ref{fig:Fig1}b). In such an experimental arrangement, the light from a fiber-coupled laser diode at 792~nm wavelength, reflected by a cube beamsplitter (Thorlabs Inc.), is focused by a 0.85~NA objective (Nikon) to a diffraction-limited spot of about 1~$\mu$m in diameter at the sample plane. The emission from the Er$^{2+}$ ions, collected through the same objective in epi-reflection geometry, is sent to a set of long-pass filters (1300 and 1350~nm cut-off wavelengths, Thorlabs Inc.) for rejecting the excitation reflected light. The filtered emission is then sent into a home-made spectrometer, which is composed simply by a rotating Pelin-Broca prism and a InP/InGaAs single photon avalanche diode (SPAD) and whose working principle is described in detail elsewhere \cite{ACSphotonics}. Such an experimental setup allows recording fluorescence maps with a diffraction limited spatial resolution of about 1.2~$\mu$m together and emission spectra with a spectral resolution of about 10-15~nm. Figure~1c shows the reflectivity map recorded at the excitation wavelength on the corner of one of the pads, selected for the investigation to include also a non-implanted silicon region for control. 

\begin{figure}[ht!]
\centering
\mbox{\includegraphics[width=7.5 cm]{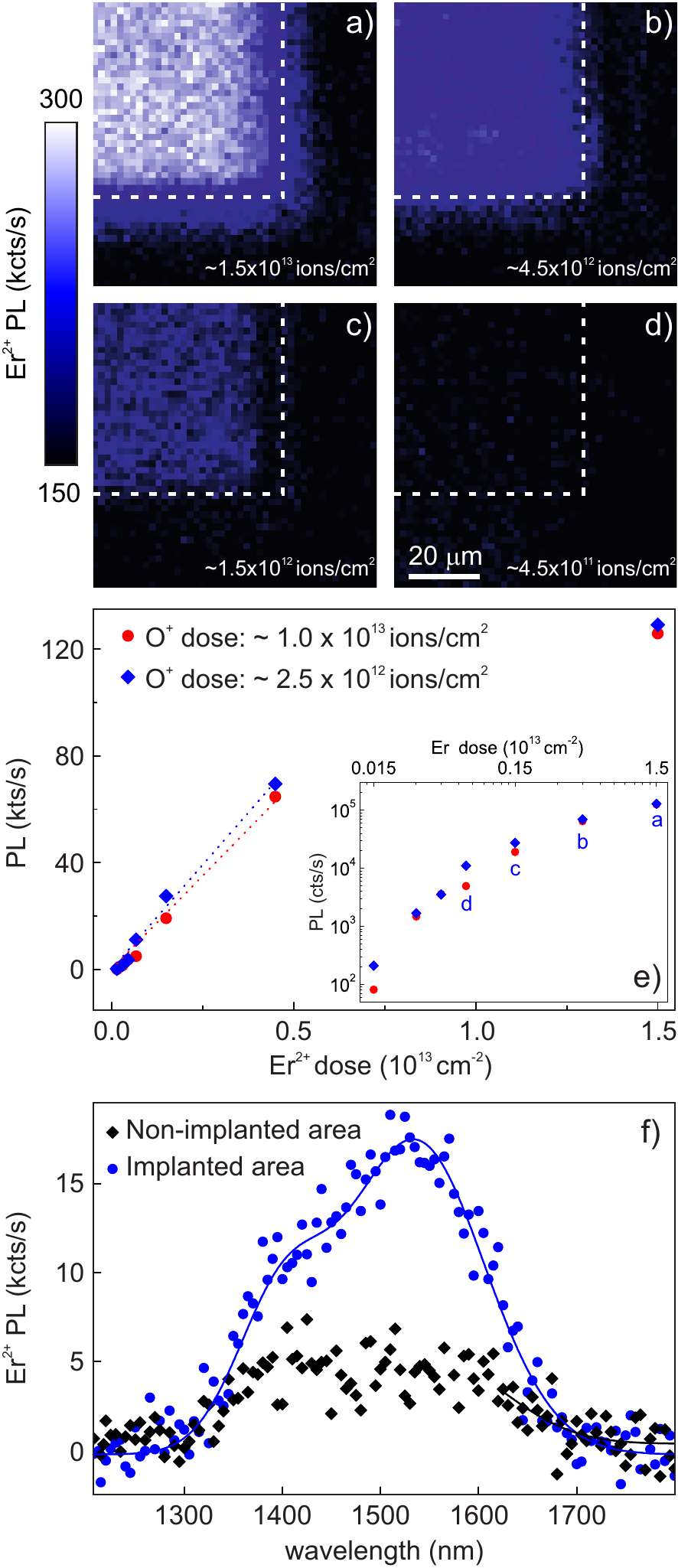}}
\caption{a)-d) Er Photoluminescence (PL) maps obtained at the same sample location as that of Fig.~\ref{fig:Fig1}c by collecting the emission above 1350~nm at different Er$^2+$ doses. The dashed lines indicate implanted areas. e) PL emission intensity recorded by the SPAD. Dashed lines represent the linear fit at low implantation doses. Inset: Log-Log plot to highlight data at the low dose. Letters refer to panels a-d. f) Spectra recorded outside (black) and inside (blue) the implanted areas. A double Gaussian fit to the data (blue line) indicates two peaks centered at 1540 and 1400~nm, respectively.}
\label{fig:Fig3}
\end{figure}

We collected PL maps on pads with effective Er$^{+2}$ ions concentrations ranging from ${1.5 \times 10^{13}}$~cm$^{-2}$ down to ${1.5 \times 10^{11}}$~cm$^{-2}$ and exposed to two different O$^{+}$ doses which ensure best emission performances. Emission maps collected on the corner of the square patches featuring higher ion implantation and higher oxygen dose are shown in Fig.~\ref{fig:Fig3}a-d. The maps also reveal homogenous implantation of the ions in each pad, at least within the sensitivity of our optical technique. The maps sketched in Fig.~\ref{fig:Fig3} are obtained by averaging 20-30 subsequent maps with an integration time of 100~ms per pixel, resulting in a typical integration time of about 3~s per pixel. Each pixel in the maps returns the number of photons recorded per second by the SPAD unit corrected by the dead-time of the detector. This procedure allows to account for the number of photons that, although reaching the sensing device, are lost due to the long dead-time of the detector (for details see Ref.~\cite{ACSphotonics}). 
By averaging the value of the signal in the patch and subsequently subtracting the background outside we have reconstructed the emission efficiency dependence on the concentration, as plotted in Fig.~\ref{fig:Fig3}e. The emission, here plotted versus the effective concentration normalized according to the out-diffusion evaluated by SIMS measurements, shows a linear growth of the efficiency with increasing ion concentration that reaches saturation roughly above $0.5 \times 10^{13}$ Er$^{2+}$/cm$^{2}$. Furthermore, both the linear and the logarithmic plots shows no significant difference between the two oxygen concentration, indicating that the annealing process optimizes the Er$^{2+}$/O$^{+}$ ratio by forming the clusters. We have also acquired emission spectra of the samples by averaging onto different locations both inside and outside the implanted area (blue dots and black diamonds respectively in Fig.~\ref{fig:Fig3}f). The spectrum acquired on the implanted region indicates the presence of a spectral feature around 1540~nm, which is typical of the erbium emission, together with a shoulder around 1400~nm. The background PL in the non-implanted area, which is visible between 1300 and 1700~nm, is attributed to residual photo-excited luminescence from the substrate.

\begin{figure}[t!]
\centering
\mbox{\includegraphics[width=12 cm]{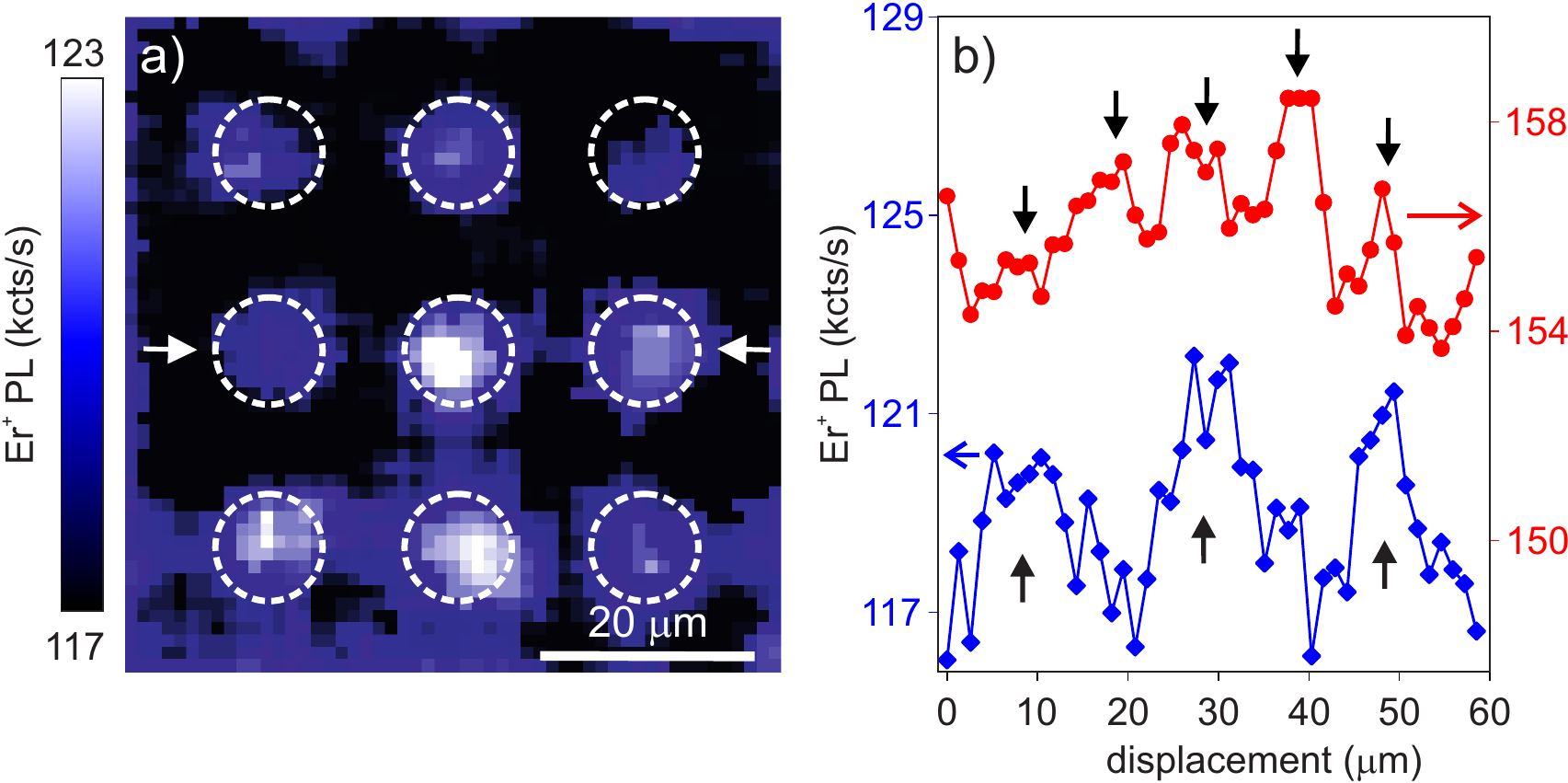}}
\caption{ a) PL map of a ${3 \times 3}$ array of 10~$\mu$m-diameter disks with Er implantation (array pitch is 20~$\mu$m). Each disk is identified by a white dashed circle as a guide to the eye. b) Intensity profile of the mid row in the array (blue) and intensity profile extracted from a ${5 \times 5}$ array of 5~$\mu$m-diameter disks with a pitch of about 10~$\mu$m. The black arrows indicate the positions of the disk centers, confirming the 20 and 10~$\mu$m pitch, respectively. In both profile the central disks show higher intensity attributed to the shape of the ion beam size in our experiment.}
\label{fig:Fig4}
\end{figure}

As a proof of concept on the potential of our implantation technique we also realized micron-scale spots featuring the higher concentration of erbium. In particular, instead of removing the PMMA by EBL from the whole pad we have just removed a set of circular disks with variable diameter. Figure~\ref{fig:Fig4}a shows the emission map collected on an area where we implanted a ${3 \times 3}$ array of 10~$\mu$m-diameter disks, placed at center-to-center distance of 20~$\mu$m (Er: ${1 \times 10^{14}}$~cm$^{-2}$ and O: ${2.5 \times 10^{13}}$~cm$^{-2}$).
Figure~\ref{fig:Fig4}b shows the profile extracted by the central row in Fig.~\ref{fig:Fig4}a (below) together with the profile extracted by a ${5 \times 5}$ array of of 5~$\mu$m-diameter disks located at a center-to-center distance of 10~$\mu$m (Er: $3 \times 10^{13}$cm$^{-2}$ and O: ${2.5 \times 10^{13}}$~cm$^{-2}$). Figure~\ref{fig:Fig4} highlights micrometer-sized isolated areas and it evidences also that the central disk features higher emission in both cases. This can be explained by the convolution between the mask and the shape of the ion implantation beam. It is also worth observing that, despite the low signal to ratio of these last measurements, the resolution appears to be mostly limited at this stage by the fabrication process rather than by the optical system, which is about 1.3~$\mu$m \cite{ACSphotonics}.

We have evaluated the lower-bound number of ions that can be detected in these conditions by combining the estimation of the overall emitted photons by the implanted area from PL measurements, with given values of the Er emission cross-section and normalized thanks to the data from SIMS measurements. 
To extract the exact number of photons emitted by the centers embedded in the illuminated area, we need to first estimate the amount of radiation directed in the semi-space of the collection optics. To do so, we have calculated, by using finite-difference time-domain (FDTD) simulations, the emission pattern of a randomly-oriented dipole emitting at 1550~nm located inside Si, placed 10~nm underneath the Si/air interface. According to these calculations only 1 over 50 photons is emitted towards the collection objective. The objective collects about 90$\%$ of the light and transmits about 30$\%$ of it in this wavelength range. By accounting for (i) the throughput of the other collection optics (about 20$\%$), (ii) the detector quantum yield (about 30$\%$) and (iii) the light lost due to the overfilling of the detector area (about 15$\%$) the overall throughput of the detection system can be estimated to be about ${5 \times 10^{-5}}$. Under the hypothesis that all the implanted ions are active, the effective emitted photon flux for an ion concentration of about ${1.5 \times 10^{12}}$~cm$^{-2}$, i.e. where the regime is still linear (see dot (c) in Fig.~\ref{fig:Fig3}e), can be estimated to be about $4 \times 10^{8}$ photons/s.

In addition, dividing the incident power at the ion plane (about 6~mW) by the excitation beam size at the focal plane (10$^{-8}$~cm$^{2}$, given a beam diameter of about 1.12~$\mu$m), one can easily obtain an excitation intensity of about ${6 \times 10^{5}}$~W/cm$^{2}$. As the emitted photon flux is equivalent to a power level of about 50~pW, the overall emission cross-section for the ions confined within the excitation volume can be estimated to be about ${8 \times 10^{-17}}$~cm$^{2}$. By assuming a typical value for the emission cross-section of individual Er$^{2+}$ of about ${5 \times 10^{-22}}$~cm$^{2}$ \cite{kippenberg2006demonstration}, we would calculate an effective number of excited ions at this nominal concentration that is about 10$^{5}$. This implies that, under the assumption that all the implanted Er atoms participate to the effect, in this study the minimum number of ions detected is about 10$^{4}$ for acquisition times that remains below 60~s. The lower-bound value for detectability is set in our measurements by two main limitations. On the one hand, the relatively high refractive index of the substrate guides the emitted light inside Si and away from the detection path, for a total loss of 95\% of the photons. On the other hand, the detector noise and excessive deadtimes limits the signal-to-noise of our measurements.

By considering the overall estimated photon flux and the estimated number of photons one can calculate a photon flux per ion of about ${4 \times 10^{3}}$ photons/s, corresponding to a PL lifetime in the millisecond range that is in line with values reported in literature. This implies a radiative lifetime for the ion in the silicon matrix of the order of the millisecond, which is very close to what reported in literature for this system.

To conclude, we demonstrated room temperature PL by low dose doping of ErO$_x$ centers implanted in natural Si at $\lambda=~1.54$~$\mu m$ via the optical transition Er$^{3+}$: $^4\mathrm{I}_{13/2}\rightarrow~^4\mathrm{I}_{15/2}$. As the implantation depth is relatively close to the top surface, we demonstrate that - after taking into account the out-diffusion from the surface due to unavoidable annealing process - the number of centers is about one order of magnitude less. The signal to noise ratio of our system allowed detecting down to about 10$^4$ Er centers at room temperature. Our estimate is supported by the derived photon flux rate per individual ion, ${\approx 4 \times 10^{3}}$ photons/s, which is close to the values reported in literature. 
The detection limit of our study is mainly set by the detector noise and the radiation losses in the silicon substrate. Hence, by engineering the substrate for effectively beaming the emitted light and employing a detector with higher performances the limit of few ion interrogation is within reach. These results set a step forward towards the employment of ErO$_x$ centers in natural silicon optoelectronic devices for quantum networks at the telecomm wavelengths. 

E.P. acknowledges JSPS, the Short Term Mobility Program 2015 of CNR and the Short Term Mobility Program 2016 of CNR.
This work was supported by a Grant-in-Aid for Basic Research (B) (25289109) from MEXT and Young Scientists (A) (15H05413) from MEXT.
The authors thank Maurizio Ferrari (CNR-IFN Trento), Giovanni Pellegrini (Politecnico di Milano) and Francesco Priolo (Università di Catania)for useful discussions.

% Bibliography
\bibliography{sample}

\clearpage

%\bibliographyfullrefs{sample}

\end{document}